# Parametric Optimization of Low Thrust Orbital Maneuvers

**Antonio Fernando Bertachini de Almeida Prado**

Instituto Nacional de Pesquisas Espaciais – INPE, São José dos Campos - SP - 12227-010 – Brazil, antonio.prado@inpe.br

## ABSTRACT

The goal of the present paper is to make a numerical analysis of parametric optimization of low thrust orbital maneuver. An orbital maneuver occurs when it is necessary to modify the orbit a space vehicle to change its function or to correct effects of perturbations. A parametric optimization is made when the thrust is not free to point to any direction, but has to follow some prescribed law, like a linear or quadratic relation with time. In this case, the optimization searches for the best value of the parameters that specify the control law and the instants to start and end each burning arc. The number of arcs can be varied, but is fixed for each optimization search. It implies the use of a suboptimal control. An optimal control gives best results, in terms of minimizing the fuel consumed, but it implies in changing the direction of the thrust and the instants to turn it on and off at every instant of time, which makes the implementation of the hardware much more difficult. Besides that, the literature shows that the differences in fuel consumptions are not large and decreases when including more thrusting arcs. A point to be considered is that, when increasing the number of arcs, the duration of the maneuver increases, and perturbations like the flattening of the Earth and third-body perturbation coming from the Sun and the Moon affects the dynamics, so the fuel consumption. This is the main goal of the present paper, which extends the technique presented here to long transfer durations under this more accurate technique. The results even show that, by choosing the appropriate times and directions to apply the thrust, the fuel consumption can be lower than the equivalent ones obtained by using a dynamics without perturbations.

## INTRODUCTION

In space engineering, an orbital maneuver is defined when we change the orbit of a spacecraft. This is important in situations where it is necessary to modify the application of a spacecraft or to perform orbital corrections in a spacecraft that had its orbital elements perturbed by forces and it is necessary to return them to their nominal values to keep the spacecraft operating. Another reason to make orbital maneuvers is to place the spacecraft in its proper orbit after the launch, which usually is not placed exactly at its nominal final orbit by the launcher.

Sometimes larger changes are required, in particular for satellites in high altitude, like geosynchronous satellites. Since fuel is scarce in space, to find orbital transfers that minimize these costs is crucial to increase the duration of a mission and its scientific and economical return.

Orbital maneuvers can be made using impulsive (Bender, 1962; Eckel, 1963, 1982; Eckel & Vinh, 1984; Santos et al., 2018; Bonasera et al, 2021) or low thrust (Tsien, 1953). Impulsive are of simple calculations and faster in implementation, but requires more fuel expenditure. Low thrust takes more time, sometimes thousands of revolutions, but are much more economical in terms of fuel expenditure.

The optimization problem can be defined has "to change the initial estate of a spacecraft (position, velocity and mass) from $\underline{r}_0$, $\underline{v}_0$ and $m_0$ at the time $t_0$, to the final state $\underline{r}_f$, $\underline{v}_f$ and $m_f$ at the time $t_f$ ($t_f \geq t_0$) using the minimum possible amount of fuel, which can be expressed by ($m_f - m_0$)", because it is assumed that the variation in mass is only due to the fuel expenditure. Figure 1 shows this transfer.

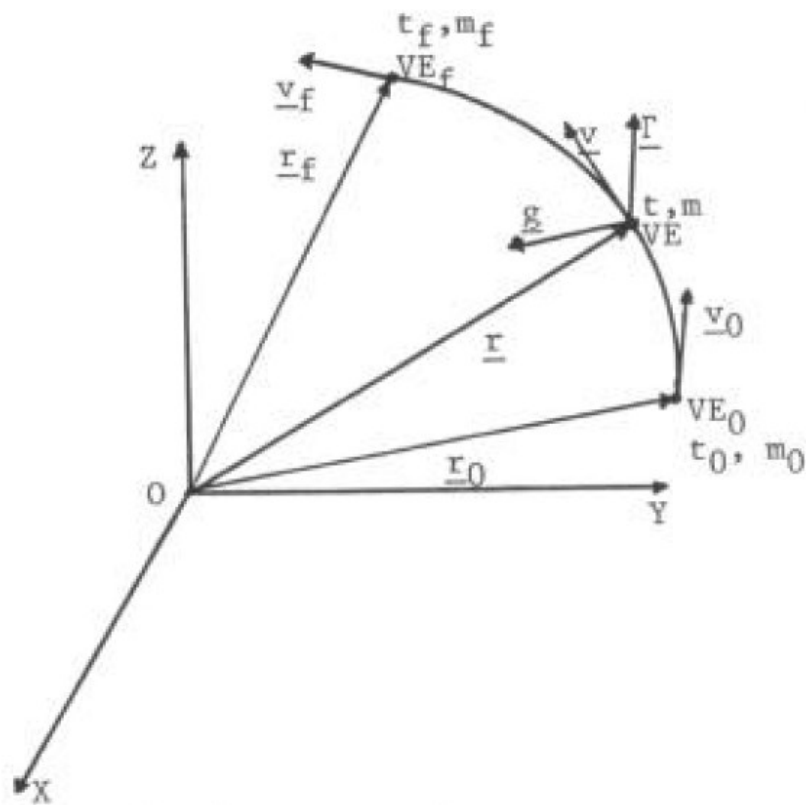

Figure 1 – Orbital transfer.

Considering optimal control (Bryson & Ho, 1975), we can choose the direction, sense and magnitude of the thrust to be applied at every instant of time of the maneuver. For the suboptimal maneuver, the direction of the thrust is not free and must follow a prescribed law. It has the advantages already explained.

Therefore, the present paper optimizes the version of the orbital transfer maneuver that has a constant magnitude for the thrust and a direction of burning that follows a linear relation with an independent variable that replaces time and describe the position of the spacecraft at a given instant of time. It is a sequence of previous researches, by adding a more accurate dynamics to allow longer transfer times.

## MATHEMATICAL MODEL

The dynamics of the spacecraft is assumed to be composed by the Keplerian term of the gravity field of the Earth, the term representing the flattening of the Earth ($J_2$) and the third-body perturbation coming from the Moon and the Sun. This thrust is assumed to have constant ejection velocity, a fixed magnitude and direction that follow a linear relation with an angular variable that defines the position of the spacecraft in its orbit. This angular variable is the true anomaly of the satellite. The number of propulsion arcs is decided before the optimization process and is an external variable to be optimized. Basically, simulations are made for different values of the number of propulsion arcs and are compared. The solution will be the instants to turn on and off the thrust and the constants that define one linear relation for the direction of burn for each propulsion arc. The function to be maximized is the final mass of the spacecraft, because by doing so we minimize that variation of mass, which is basically the fuel consumed.

The direction of burning can be described by linear relations like Equs. (1) and (2) shown below, where the planar angle of pitch is ($A$) and the out-of-plane yaw angle is ($B$).

$$A = A_0 + A' * ( x - x_S ) \qquad (1)$$
$$B = B_0 + B' * ( x - x_S ) \qquad (2)$$

with $A_0$, $B_0$, $A'$, $B'$ constants to be found by the optimization method, $x$ the independent angular variable describing the position of the spacecraft and $x_S$ representing the instant the thrust is turned-on in a particular arc. There is one set of equations like this one for each propulsion arc assumed for the maneuver. It means that there are six variables to be optimized for each "propulsion arc" used.

The equations of motion of the spacecraft are:

$$\ddot{x} = -Gm_1 \frac{(x-x_1)}{r_1^3} - Gm_2 \frac{(x-x_2)}{r_2^3} - Gm_3 \frac{(x-x_3)}{r_3^3} - Gm_1 J_2 r_{M1}^2 \left(\frac{3x}{2r_1^5}\right) \quad (3)$$

$$\ddot{y} = -Gm_1 \frac{(y-y_1)}{r_1^3} - Gm_2 \frac{(y-y_2)}{r_2^3} - Gm_3 \frac{(y-yx_3)}{r_3^3} - Gm_1 J_2 r_{M1}^2 \left(\frac{3y}{2r_1^5}\right) \quad (4)$$

$$\ddot{z} = -Gm_1 \frac{(z-z_1)}{r_1^3} - Gm_2 \frac{(z-z_2)}{r_2^3} - -Gm_3 \frac{(z-z_3)}{r_3^3} Gm_1 J_2 r_{M1}^2 \left(\frac{3z}{2r_1^5}\right) \quad (5)$$

Where (x,y,z) are the coordinates of the spacecraft; $(x_1,y_1,z_1)$ the coordinates of the Earth; $(x_2,y_2,z_2)$ the coordinates of the Sun; $(x_3,y_3,z_3)$ the coordinates of the Moon; G is the universal gravitational constant; $r_1$ is the distance from the spacecraft to $M_1$ (Earth); $r_2$ is the distance from the spacecraft to $M_2$ (Sun); $r_3$ is the distance from the spacecraft to $M_3$ (Moon); the radius of Earth is $r_{M1}$; $J_2$ represents the flattening of the Earth and $m_1$, $m_2$, $m_3$ the masses of the Earth, Sun and Moon, respectively.

## NUMERICAL METHOD

After setting up the mathematical model, it is necessary to choose a technique to solve the nonlinear programming problem that we have. The method selected is the gradient projection method (Bazarra & Sheetty, 1979; Luemberger, 1973; Prado & Rios-Neto, 1989; Prado. 2001). The steps are described next. After the numerical integration, we:

i) First force the system to satisfy the constraints, by following the sequence:

$$\boldsymbol{u}_{i+1} = \boldsymbol{u}_i - \nabla \mathbf{f}^T \cdot \left[\nabla \mathbf{f} \cdot \nabla \mathbf{f}^T\right]^{-1} \mathbf{f} \quad (6)$$

where f is the vector of active constraints;

ii) After that we try to minimize the fuel consumed by using the sequence.
iii)

$$\boldsymbol{u}_{i+1} = \boldsymbol{u}_i + \alpha \frac{\mathbf{d}}{|\mathbf{d}|} \quad (7)$$

with:

$$\alpha = \gamma \frac{J(\boldsymbol{u})}{\nabla J(\boldsymbol{u}) \cdot \mathbf{d}} \quad (8)$$

$$\mathbf{d} = -\left(\mathbf{I} - \nabla \mathbf{f}^T \left[\nabla \mathbf{f} \cdot \nabla \mathbf{f}^T\right]^{-1} \mathbf{f}\right) \nabla J(\boldsymbol{u}) \quad (9)$$

where **I** the unit matrix, **d** is the direction of the search, J is the fuel consumed and γ a parameter that we choose to control size step.

We follow this sequence until $|u_{i+1} - u_i| < \varepsilon$ in Equs. (6) and (7), where ε is a specified tolerance for the iterations. Different values are used for this tolerance to test the system.

## RESULTS

The results consist of simulating suboptimal maneuvers with different number of propulsion arcs and durations using the numerical algorithm just described, with the goal of finding the complete solutions of the transfer: the instants to turn on and off the propulsion, the variables that define the linear direction of the thrust, the fuel consumptions and number of revolutions made by the spacecraft during the maneuver. We tried maneuvers with 1, 2, 4 and 8 propulsion arcs. Table 1 shows the results.

Table 1

Orbital maneuvers with 1, 2, 4 and 8 "propulsion arcs"

| Arc | $x_S$(deg) | $x_e$(deg) | $A_0$(deg) | $B_0$(deg) | $A'$ | $B'$ | Fuel-kg | Number of revolutions |
|---|---|---|---|---|---|---|---|---|
| 1 | 63.9 | 5010.6 | 10.1 | 12.3 | 0.0135 | -0.002 | 7.37 | 13.9 |
| | | | | | | | | |
| 1 | 49.2 | 2548.6 | 8.3 | -20.2 | 0.015 | 0.425 | ------ | ------- |
| 2 | 3050.3 | 5567.3 | 6.1 | 22.3 | -0.136 | -0.031 | 7.11 | 15.5 |
| | | | | | | | | |
| 1 | 325.1 | 1578.2 | 5.5 | -13.2 | 0.019 | -0.047 | ------ | -------- |
| 2 | 2043.4 | 4541.9 | 15.6 | 21.1 | -0.184 | -0.270 | ------ | -------- |
| 3 | 4901.1 | 6151.1 | 9.2 | -35.5 | -0.009 | 0.487 | ------ | -------- |
| 4 | 6513.9 | 7761.9 | 3.5 | 21.2 | -0.167 | -0.233 | 6.05 | 21.6 |
| | | | | | | | | |
| 1 | 630.3 | 1254.3 | 10.1 | -12.4 | -0.011 | -0.078 | ------ | -------- |
| 2 | 1453.1 | 2061.9 | 10.9 | 14.1 | -0.198 | -0.340 | ------ | -------- |
| 3 | 2543.0 | 3044.9 | 12.2 | -15.3 | -0.028 | 0.987 | ------ | ------- |
| 4 | 3401.2 | 4030.8 | 16.1 | 13.0 | -0.076 | -0.089 | ------ | ------- |
| 5 | 4561.8 | 5188.9 | 7.1 | -9.2 | 0.035 | -0.189 | ------ | --------- |
| 6 | 5546.8 | 6170.9 | 9.1 | 21.8 | -0.238 | -0.198 | ------ | ---------- |
| 7 | 6531.1 | 7153.7 | 8.0 | -19.2 | -0.006 | 0.963 | ------ | -------- |
| 8 | 7756.6 | 8313.2 | 11.4 | 43.8 | -0.980 | -0.023 | 5.67 | 23.1 |

The results confirms some expected facts, like that the fuel consumed decreases with the increasing number of propulsion arcs. It is 7.37 kg for a single propulsion arc, 7.11 kg for two propulsion arcs, 6.05 kg for four propulsion arcs and 5.67 kg for eight propulsion arcs. This fact just confirms that the more accurate dynamics does not change this aspect.

A more interesting study is to verify the differences in the results when the perturbations coming from the Sun, the Moon and the flattening of the Earth are included in the dynamics. In this case, the system finds solution that are slight different and that can use the perturbations to reduce a little bit the fuel consumption. Table 2 summarizes the differences if the fuel consumed and the number of revolutions used to make the maneuver.

Table 2

Fuel consumed and number of revolutions for the two-body and perturbed dynamics

| Number of propulsion Arcs | Fuel-kg for the "perturbed" dynamics | Number of revolutions for the "perturbed" dynamics | Fuel-kg for the "two-body" dynamics | Number of revolutions for the "two-body" dynamics | Saving (%) |
|---|---|---|---|---|---|
| 1 | 7.11 | 13.5 | 7.37 | 13.9 | 3.53% |
| 2 | 6.84 | 15.2 | 7.11 | 15.5 | 3.80% |
| 4 | 5.85 | 20.9 | 6.05 | 21.6 | 3.31% |
| 8 | 5.41 | 22.8 | 5.67 | 23.1 | 4.59% |

The results show savings in the order of 4%, reaching a maximum of 4.59%. Those results are not small enough to be neglected and should be included in real mission's evaluations. As a parameter of comparison, Table 3 shows the savings obtained by increasing the number of propulsion arcs, which makes the transfer much longer. Comparing those results, we see that increasing the number of propulsion arcs from one to two, we have a savings in the fuel consumed of 3.53%, which is similar to the savings obtained by considering a perturbed dynamics. Basically, the solution of the parametric optimization problem searches for the best moments to apply the propulsion such that you get help from the perturbative forces. The results are equivalent of going from one to two propulsion arcs, so it is not negligible. When going from two to four propulsion arcs, the difference increases to 16.53% in fuel consumed, with an extra savings of 6.28% if we use eight propulsion arcs. Even those larger savings are about five and two times the savings provided by exploring the more accurate dynamics. Of course there is also the additional advantage that this dynamics gives results that are closer to reality, since the perturbations are there.

Table 3

Savings obtained by increasing the number of propulsion arcs.

| Number of propulsion Arcs | Saving (%) compared to 1 arc |
|---|---|
| 1 | ------- |
| 2 | 3.53% |
| 4 | 16.53% |
| 8 | 6.28% |

## CONCLUSIONS

In the present research we transformed the minimum fuel consumption spacecraft maneuver problem in a parametric optimization problem, where the goal was to find the parameters that describe the beginning and end of each propulsion arc and the parameters that define the linear law for the variation of the direction of the propulsion that minimizes fuel consumption.

For the control system, we used a propulsion that has fixed magnitude and linearly varying direction of burning. For the dynamics, we made simulations using only a Keplerian motion for the spacecraft around the Earth and a mathematical model that includes the perturbations coming from the Sun, the Moon and the flattening of the Earth.

The results show clearly that: the increasing in the number of propulsion arcs reduces the fuel consumed in both dynamics, in particupar when changing from two to four propulsion arcs; the dynamics including the perturbations can help the maneuver by decreasing the fuel consumed with savings in the order of 3-5%, which are not negligible. Those savings are obtained by choosing the best moments to turn on the propulsion system such that it adds force to the perturbation to make the desired maneuver.

## ACKNOWLEDGEMENTS

The author thanks São Paulo Research Foundation (FAPESP) for the grant 2016/24561-0 that supported the development of the present paper.
.